\documentclass[a4paper,11pt]{article}
\usepackage{pos}
\usepackage{overpic}

\newcommand{\pp}{\pi^+\pi^-}

\newcommand{\kk}{K^+K^-}

\newcommand{\EE}{e^+e^-}

\newcommand{\etap}{\eta^\prime}
\newcommand{\psp}{\psi(2S)}
\newcommand{\psip}{\psi(2S)}

\newcommand{\jpsi}{J/\psi}
\newcommand{\piz}{\pi^0}

\newcommand{\xyz}{\rm XYZ}

\newcommand{\zc}{Z_c(3900)}
\newcommand{\zcp}{Z_c(4020)}

\newcommand{\ppjpsi}{\pi^+\pi^-J/\psi}
\newcommand{\hc}{h_c}
\newcommand{\pphc}{\pi^+\pi^-\hc}

\newcommand{\ccb}{c\bar{c}}
\newcommand{\ppb}{p\bar{p}}

\title{New experimental results on light and heavy hadrons}

\ShortTitle{New experimental results ...}

\author*[a,b]{Chang-Zheng Yuan}
\affiliation[a]{Institute of High Energy Physics, Chinese Academy of
Sciences, \\
19B Yuquan Road, Beijing, China}
\affiliation[b]{University of Chinese Academy of Sciences, \\
19A Yuquan Road, Beijing, China}

\emailAdd{yuancz@ihep.ac.cn}

\abstract{
We review the experimental progress on the study of light and heavy hadrons and 
focus on recent discoveries of the $\xyz$ states, pentaquark states,
$T_{cs}$ and $T_{cc}$ states in the heavy hadron sector, and on 
the $\pi_1(1600)$ and the states close to the $\ppb$ threshold in
the light hadron sector. The observations suggest that we did 
observe hadronic molecules and we also observed hadronic states 
with some other quark configurations. 
}

\FullConference{%
  *** Particles and Nuclei International Conference - PANIC2021 ***\\
  *** 5 - 10 September, 2021 ***\\
  *** Online ***
}

\begin{document}
\maketitle

\section{Introduction}

Hadron spectroscopy is a field of frequent discoveries and surprises,
and the theoretical difficulties in understanding the strong interaction
in the color-confinement regime make the field even more fascinating. 
The tremendous data collected by the BaBar, Belle, BESIII, COMPASS, LHCb, 
and other experiments and improved theoretical tools developed to analyze the 
experimental data result in rapid progress of the field. 

In this talk, we review the progress in both heavy hadrons and light hadrons,
and we focus on those states with exotic properties, including the 
quarkoniumlike states, also known as the $\xyz$ states, pentaquark states,
and light hadrons close to thresholds or with exotic quantum numbers.

\section{New results on the heavy hadrons}

In the conventional quark model, mesons are composed of one quark
and one anti-quark, while baryons are composed of three quarks.
However, many quarkoniumlike states were discovered at two
$B$-factories BaBar and Belle~\cite{PBFB} in the first decade of 
the 21st century. Whereas some of these
are good candidates of quarkonium states, many other states have
exotic properties, which may indicate that exotic states, such as
multi-quark state, hadronic molecule, or hybrid, have been
observed~\cite{reviews}.

BaBar and Belle experiments finished their data taking in 2008 and
2010, respectively, and the data are still used for various
physics analyses. BESIII~\cite{ijmpa_review} and 
LHCb~\cite{lhcb} experiments started data taking
and contributed to the study of the $\xyz$ particles, pentaquark
states, and other heavy hadrons. Most of the discoveries of the 
such states were made at these four experiments.

\begin{figure*}[htbp]
\centering
  \includegraphics[width=0.95\textwidth]{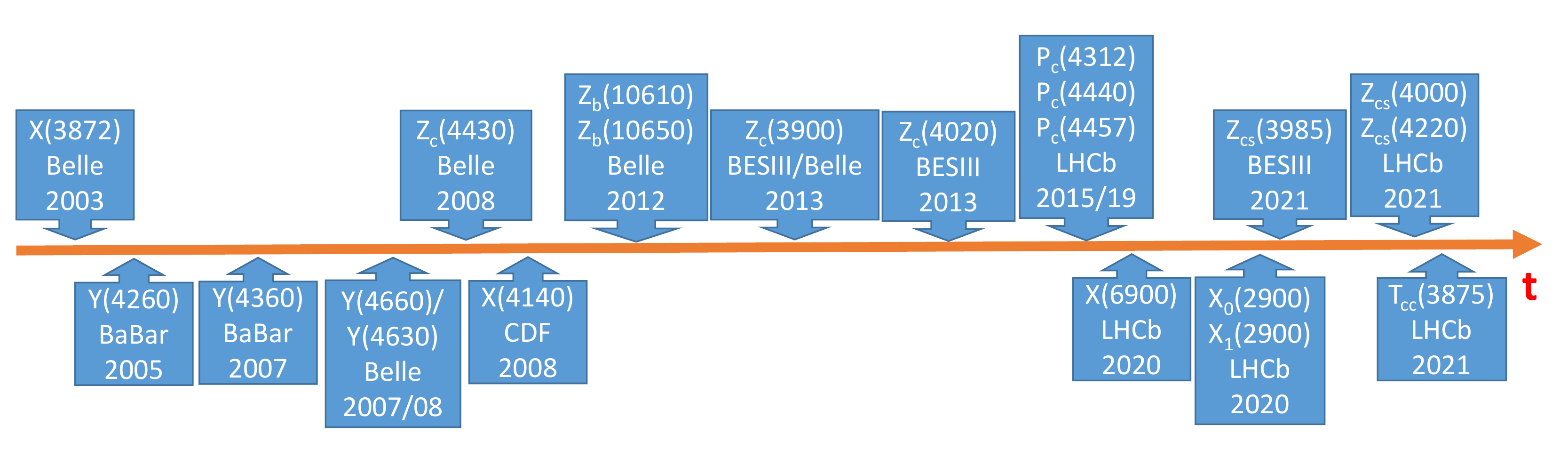}
\caption{Discovery of heavy exotic states from experiments.}\label{fig:XYZ}
\end{figure*}

Figure~\ref{fig:XYZ} shows the history of the discovery of the heavy 
exotic states, started from the observation of the $X(3872)$ 
in 2003~\cite{Bellex}. It seems to us that the new spectrum has emerged although 
more effort is need to understand the exact nature of them.
We show some recent experimental results on these particles. 

\subsection{The $X(3872)$}

The $X(3872)$ was observed in 2003 by the Belle 
experiment~\cite{Bellex}, and it was 
confirmed later by CDF~\cite{CDFx} and $D0$~\cite{D0x} experiments 
in $p\bar{p}$ collision. 
After almost 20 years' study, we know this state much better than any of the
other similar states. 

The mass of the $X(3872)$ has been measured as $3871.65\pm 0.06$~MeV~\cite{pdg}, 
which is lower than the mass threshold of $\bar{D}^0D^{*0}$, $3871.69\pm 0.11$~MeV,
by $0.04\pm 0.12$~MeV, to be compared with the bounding energy of the deuteron
of 2.2~MeV. If the $X(3872)$ is a molecule of $\bar{D}^0D^{*0}$, its size 
will be larger than 5~fm, much larger than the size of a typical hadron.

The width measurements are less precise and model dependent since 
the $X(3872)$ is very narrow and the mass resolution of the experiments
is usually much larger than the intrinsic width. Fitting the 
$\ppjpsi$ invariant mass distribution with a Breit-Wigner function, 
LHCb reported a width of about 1~MeV (the mass resolution is 2.4--3.0~MeV);
and the fit with a Flatt\'e function with constraints from other measurements
yields a FWHM of 0.22~MeV which depends strongly on 
the $X(3872)\to \bar{D}^0D^{*0}$ coupling~\cite{LHCb_x_width1,LHCb_x_width2}. 
Although the statistics are
low at BESIII experiments, the high efficiencies of reconstructing 
all the $X(3872)$ decays modes and the very good mass resolution in
the $\bar{D}^0D^{*0}$ mode ($<1$~MeV) make it possible to measure
the line shape of the $X(3872)$ state~\cite{bes3_x_brs}.

The production of the $X(3872)$ has been reported in many different 
kinds of processes, in $B$ and $B_s$ meson decays, in $\Lambda_b$
baryon decays, in $p\bar{p}$ and $pp$ collisions, and 
in $\EE$ annihilation~\cite{BES3x}. Recently, evidence for $X(3872)$ 
production in $PbPb$ and two-photon collisions is reported.

CMS experiment reported a $4.2\sigma$ signal of the $X(3872)$ in 
$PbPb$ collision at 5.02~TeV~\cite{CMS:2021znk}, and it is interesting to note that 
its production rate relative to the $\psip$ is much larger than 
in the $pp$ collision at 7 and 8~TeV, although the uncertainty is
large. If this is confirmed, this is a supplemental information 
to understand the nature of this state.

Belle experiment searched for the $X(3872)$ in $\gamma\gamma^*$
fusion~\cite{Belle:2020ndp}. Three events are observed in the signal region, corresponds 
to a statistical significance of $3.2\sigma$. Since we know that
the $X(3872)$ has $J^{PC}=1^{++}$, it cannot be produced in two real-photon 
collision, the production requires at least one of the photons is virtual.

The total production rate of the $X(3872)$ in $B$ decays was measured 
by reconstructing a $B^-$ and a charged kaon from $B^+$ decays and 
checking the recoiling mass of the $B^-K^+$ system. BaBar observed a small peak, 
corresponding to a $3.0\sigma$ significance at the $X(3872)$ signal region, 
and measured the branching fraction of $B^+\to K^+X(3872)$ as 
$(2.1\pm 0.6\pm 0.3)\times 10^{-4}$~\cite{BaBar:2019hzd}. Belle did the same analysis, 
but the signal is less significant and the resulting branching fraction 
is $(1.2\pm 1.1\pm 0.1)\times 10^{-4}$ and the signal significance 
is $1.1\sigma$~\cite{Belle:2017psv}. Although the signals are not very significant, 
we know this process must exist because this state has been observed in its 
many exclusive decays. One can use these measurements combined with 
other information, such as the product branching fractions and 
the ratio of the branching fractions, to determine the decay branching
fractions of the $X(3872)$, including its decays to open charm final states,
hadronic transitions, and radiative transitions. 
There could be a small branching fraction to light hadrons, but 
no experiment has observed any of them. 

The authors of Ref.~\cite{Li:2019kpj} did a global fit to the currently 
available experimental measurements of the product branching fractions and 
the ratios of the branching fractions. It is found that the branching fraction 
of open charm decay is around 50\% and that of each hadronic transition 
is at a few per cent level, there is still around one-third of the 
$X(3872)$ decays unknown. This should be searched for in the future 
experiments like BESIII and Belle II. 

One still very confusing decay mode is $X(3872)\to \gamma \psp$. 
There have been four different measurements. The BaBar experiment
claimed $3.5\sigma$ evidence of this mode and a production rate 
relative to  $X(3872)\to \gamma \jpsi$ is $3.4\pm 1.4$~\cite{R_in_babar}, 
but Belle failed to find significant signal and the ratio was 
measured to be less than $2.1$ at the 90\% C.L.~\cite{R_in_belle}. 
Three years later, LHCb did the same analysis and the found 
a $4.4\sigma$ signal with a ratio of $2.46\pm 0.81$~\cite{R_in_lhcb}, but a recent 
BESIII measurement found no signal and a much stringent 
upper limit of the ratio is determined to be 0.59 at the 90\% C.L.~\cite{bes3_x_brs}. 
So we have four experiments here, two claimed evidence and the 
other two observed nothing. So it is still not clear 
whether this channel, $X(3872)\to \gamma \psp$, exists or 
if it exists, how small the branching fraction is. 

\subsection{The $Y$ states}

The $Y$ states were discovered in the initial state radiation in 
the $B$-factory experiments, and they have $J^{PC}=1^{--}$. 
So these state can also be produced directly in $\EE$ annihilation
experiment like BESIII. In this case, much larger statistics 
are achieved and these states, the $Y(4260)$, $Y(4360)$, $Y(4660)$, 
and so on, are measured with improved precision. 

The $Y(4260)$ was observed in 2005 by BaBar experiment~\cite{babar_y4260} and the most 
precise measurement is from the BESIII experiment~\cite{BESIII:2016bnd}. By doing a high
luminosity energy scan in the vicinity of the $Y(4260)$, 
BESIII found the peak of the $Y(4260)$ is much lower (so now named
the $Y(4220)$) than that from
previous measurements and the width is narrow, and there is a high
mass shoulder with a mass of 4.32~GeV if fitted with a BW 
function. Since then, more new decay modes of the $Y(4220)$ were 
observed including $\pphc$, $\omega\chi_{c0}$, and
$\pi \bar{D} D^*+c.c.$  

A global fit~\cite{Gao:2017sqa} to these four modes was performed, and 
One determines the mass of the $Y(4220)$ as $4220\pm 6$~MeV 
and the width of $56\pm 8$~MeV. It is interesting to point 
out that the mass of this resonance is quite close to the 
threshold of threshold of $D_s^{*+}D_s^{*-}$ which is $4224$~MeV.
Since there are more $Y(4220)$ decay modes observed 
($\pp\psp$, $\eta_c\pp\piz$, and so on), 
this combined fit can be updated with more information.

For the state at 4.66~GeV, it was observed in $\EE\to \pp\psp$ by Belle~\cite{belle_y4660},
and confirmed by BaBar~\cite{BaBar:2012hpr}. The peak position is at around 4.66~GeV, 
thus it is called the $Y(4660)$. There is another state observed 
in $\EE\to \Lambda_c \bar{\Lambda}_c$ 
by the Belle experiment~\cite{Belle:2008xmh}, but the peak is at 
around 4.63~GeV, although the error is large. It is not clear
whether these two states are the same or whether there are two states 
in this energy region. 

BESIII data on $\EE\to \pp\psp$ mode from 4 to 4.7~GeV confirmed 
the Belle and BaBar observations with much improved precision~\cite{BESIII:2021njb},
BESIII has data now covering from threshold to 4.95~GeV, 
comparable precision as at 4.6~GeV is expected at high energies, so we expect
better measurement of the $Y(4660)$ state from BESIII soon.

Belle reported measurements of two open-charm final states. 
There is a very beautiful peak observed at around 4.63~GeV 
in $D_s^+D_{s1}(2536)^-+c.c.$ mode and the signal significance 
is $5.9\sigma$~\cite{Belle:2019qoi}. The signal in $D_s^+D_{s2}(2573)^-+c.c.$ mode 
is not so significant, is only $3.4\sigma$~\cite{Belle:2020wtd}. 

If we put all these information together, we can find that
the peak position is about 4.65~GeV in $\pp\psp$ mode, and
that in open charm baryon and meson pair final states is 
below 4.65~GeV, There are differences from different final states.
We need more information to really understand the structures 
in this mass region. 

\subsection{Charged quarkonium states}

These include the $Z_c$, $Z_b$, and also the $Z_{cs}$ states.
Since these states decay into final states with one pair of 
heavy quarks and charged, there must be at least four quarks
in their configuration. 

The $Z_c(3900)$ discovered by BESIII~\cite{zc3900} and 
Belle~\cite{Belle_zc} is quite close to the
$\bar{D}D^*$ threshold, and the $Z_c(4020)$ discovered by BESIII is 
quite close to the $\bar{D}^*D^*$ threshold~\cite{BESIII:2013ouc}, 
while the $Z_c(4430)$ discovered by the Belle~\cite{Belle_zc4430} 
is not quite close to any of the open charm
threshold. In the bottom sector, The $Z_b(10610)$ and $Z_b(10650)$ 
discovered by Belle~\cite{Belle:2011aa} are close to the $\bar{B}B^*$ and $\bar{B}^*B^*$ 
thresholds, respectively. 

These states have been observed for some time. Recent studies try to 
search for states with one of the four quarks replaced by a different
quark, for example, the $Z_{cs}$ states with quark content $\ccb u \bar{s}$.
There are three different measurements, one from Belle in 
$\EE\to\kk\jpsi$~\cite{Belle:2014fgf},
another from BESIII in $\EE\to K(DD_s^*+D^*D_s)$~\cite{BESIII:2020qkh}, 
and the third from LHCb in $B^+\to\jpsi K^+ \phi$~\cite{LHCb:2021uow}. 

No significant signal was observed in Belle data due to low statistics 
of the process produced via ISR. BESIII announced observation of a 
near-threshold structure $Z_{cs}(3985)$ in the $K^+$ recoil-mass spectra 
in $\EE\to K^+(D^-_sD^{*0}+D^{*-}_sD^{0})$~\cite{BESIII:2020qkh} with a mass of 
3983~MeV and a width of about 10~MeV; and LHCb reported two 
resonances decaying into $K^\pm\jpsi$, the $Z_{cs}(4000)$ with a mass
of 4003~MeV and a width of about 131~MeV, and the $Z_{cs}(4220)$ 
with a mass of 4216~MeV and a width of about 233~MeV~\cite{LHCb:2021uow}.
The widths of the $Z_{cs}(3985)$ and $Z_{cs}(4000)$ are quite different,
so they could not be the same state. Maybe one of them is the strange
partner of the $Z_c(3900)$ with the $d$ quark replaced with an $s$
quark.

\subsection{Pentaquark states}

In the decay of $\Lambda_b \to \jpsi p K^-$ analyzed by the LHCb experiments,
there are three very narrow peaks in the invariant mass distribution of
$\jpsi p$~\cite{LHCb:2019kea}. In a simple fit to the invariant mass spectrum, the resonance 
parameters of the $P_c(4312)^+$, $P_c(4440)^+$, and $P_c(4457)^+$ are 
determined. They are all narrow, and the $P_c(4312)^+$ state peaks right 
below the $\Sigma_c^+\bar{D}^0$ threshold, the $P_c(4457)^+$ state 
peaks right below the $\Sigma_c^+\bar{D}^{*0}$ threshold, 
while the $P_c(4457)^+$ state peaks about 20~MeV below it.
Since the statistics are large and these structures are quite narrow,
partial wave analysis (PWA) to the data sample is still not available,
and the spin-parities are not known yet. 

Being so close to the thresholds, they are very good candidates 
for the molecules of a charmed baryon and an anti-charmed meson,
and we expect more similar states close to the other baryon-meson 
thresholds.

\subsection{States with four flavors ($T_{cs}$) or two heavy quarks ($T_{cc}$)}

States with four different flavors have been search for and evidence 
($3.9\sigma$) for two states ($X_0(2900)$ and $X_1(2900)$) in $D^-K^+$ 
system was reported from a PWA of $B^+\to D^+D^-K^+$ events by
the LHCb experiment~\cite{LHCb:2020bls}. 
These states, if confirmed, will be manifestly exotic, charm-strange 
$T_{cs}$ states with quark content $\bar{c}du\bar{s}$.

LHCb also reported new observation of a state with two charmed quarks,
$T_{cc}(3875)$ with a mass a few hundred $k$eV lower than the $D^0D^{*+}$ threshold
and a width much less than 1~MeV~\cite{LHCb:2021vvq}. 
If it is interpreted as a molecule of 
two charmed mesons, its size will be around 7.5~fm, pretty large than a
normal hadrons of 1~fm.

\subsection{Summary on heavy hadrons}

If we summarize these heavy hadrons, what we find is that some of them 
are quite close to the thresholds of two heavy objects, either two heavy 
flavor mesons or one heavy flavor meson and one heavy flavor baryon, 
like the $X(3872)$ ($\bar{D}^0D^{*0}$), 
$Y(4220)$ ($D_s^{*+}D_s^{*-}$ or $\bar{D}D_1)$,
$\zc^+$ ($\bar{D}^0D^{*+}$),
$\zcp^+$ ($\bar{D}^{*0}D^{*+}$),
$Z_{cs}^+$ ($\bar{D}^0D_s^{*+}$),
$Z_b(10610)^+$ ($\bar{B}^0B^{*+}$),
$Z_b(10650)^+$ ($\bar{B}^{*0}B^{*+}$)
$P_c(4312)^+$ ($\Sigma_c^+\bar{D}^0$),
$P_c(4457)^+$ and $P_c(4457)^+$ ($\Sigma_c^+\bar{D}^{*0}$), and
$T_{cc}(3875)^+$ ($D^0D^{*+}$); and some other states
are not close to such thresholds, such as the $Y(4360)$
$Y(4660)$, $Z_c(4430)^+$, $Z_{cs}(4220)^+$, and $X(6900)$.

These may suggest that we did observe the hadronic molecules 
close to thresholds and we also observed hadronic states 
with some other quark configurations like compact tetraquark 
states and so on. At the same time, these suggest that
similar dynamics will also play important role in light 
quark sector, and this makes the understanding of the light
hadron spectroscopy even harder. 

\section{Light hadron spectroscopy}

We present two selected topics on light hadrons, the $\pi_1(1600)$
with exotic quantum numbers and the states close to $\ppb$ threshold.

\subsection{The $\pi_1(1600)$ from COMPASS}

The $\pi_1(1600)$ is a state with $J^{PC}=1^{-+}$, quantum numbers cannot
reach by a $q\bar{q}$ state. A recent COMPASS 
analysis~\cite{COMPASS:2021ogp} tries to understand 
the tensions observed between COMPASS and previous experiments, 
including the BNL E852 experiment, VES experiment, and also 
COMPASS analysis with less waves in the PWA. 

It is concluded that: 
1) conflicting conclusions from previous 
studies can be attributed to different models and treatment 
of $t^\prime$-dependence of the amplitudes; 
2) Deck model can describe data in spectral shape and 
$t^\prime$-dependence for the $J^{PC}=1^{-+}$ and other amplitudes;
3) Freed-isobar fit results indicate that the $P$-wave $\pp$-amplitude 
is dominated by $\rho(770)$ for both the $\pi_1(1600)$ and the 
non-resonant $J^{PC}=1^{-+}$ amplitudes.

The final conclusion quoted from the paper is shown here:
"These findings largely confirm the underlying assumptions 
for the isobar model used in all previous partial-wave 
analyses addressing the $J^{PC}=1^{-+}$ amplitude.” 
Or simply speaking, the $\pi_1(1600)$ really exists. 

\subsection{States close to $\ppb$ threshold in $\jpsi$ decays}

BESIII studies of the $X(1835)\to \etap\pp$ line shape
with 1.3 billion $\jpsi$ events revealed an anomalous structure in
its line shape at the $\ppb$ threshold that could be equally well
described as interference with a new narrow resonance that has a
mass nearly equal to $2m_p$ or a wide resonance with an
anomalously strong coupling to the $\ppb$ final
state~\cite{Ablikim:2016itz}.

States in this mass region were also reported in
$\jpsi\to\omega\eta\pp$~\cite{Ablikim:2011pu}, $\gamma
3(\pp)$~\cite{Ablikim:2013spp}, and
$\gamma\gamma\phi$~\cite{Ablikim:2018hxj}, 
but not in $\jpsi\to\omega\etap\pp$~\cite{BESIII:2019sfz}. 

\section{Summary and Perspectives}

There are a lot of progress in the experimental study of hadron spectroscopy.
Spectroscopy of hadronic molecules need to be further investigated, and
states formed by other dynamics may have been discovered too.
Study of similar states in the light quark sector ($u$, $d$, $s$) may 
reveal even richer phenomena of strong interaction. It is expected that
more results will be produced by the Belle II, BESIII, COMPASS, GlueX, 
and LHCb experiments.

\section*{Acknowledgments}

I thank the organizers for inviting me to give this review talk. 
This work is supported in part
by National Key Research and Development Program of China 
(No.~2020YFA0406300), and National Natural Science Foundation
of China (NSFC, Nos. 11961141012, 11835012, and
11521505).

\end{document}